\documentclass[%
twocolumn,
 amsmath,amssymb,
 aps,
 pra,
]{revtex4-1}

\usepackage{graphicx}
\usepackage{dcolumn}
\usepackage{bm}

\begin{document}




\title{Comment on ``Measuring the enhancement factor of the hyperpolarized Xe in nuclear magnetic resonance gyroscopes"}
\date{\today }
\author{B. Saam}
\email{brian.saam@wsu.edu}
\affiliation{Department of Physics and Astronomy, Washington State University, P.O. Box 642814, Pullman, WA 99164-2814, USA}

\date{\today}


\maketitle


In \cite{Xu2021},  the authors propose several erroneous ideas regarding the nature of the enhancement factor $\kappa$ and report an experimental result for it that is more than an order of magnitude greater than theoretical estimates \cite{Walker1989} and previous experimental results \cite{Fang2014}. Moreover, there is no plausible physical basis or discussion of how their experiment might have generated such a large discrepancy.

The authors properly define the effective magnetic field seen by alkali-metal atoms due to polarized noble-gas atoms, including the enhancement factor $\kappa$ in their Eq.~(1), and the goal of the work is clearly to measure $\kappa$ for the Cs-Xe pair. However, they repeatedly confuse enhancement of the field sensed by Cs atoms due to the Fermi-contact magnetic field produced by polarized $^{129}$Xe atoms (i.e., the definition of $\kappa$), and enhancement of the $^{129}$Xe nuclear polarization produced by SEOP. They appear to be unaware of the measurement of $\kappa\approx 700$ for Cs-Xe made by Fang, et al.~\cite{Fang2014} from 2014 and do not cite it. They make the further unphysical argument that their much larger value results from making their measurement in a low applied magnetic field of 2.5~$\mu$T; indeed, the measurement by Fang, et al.~\cite{Fang2014} was made at an even lower applied field. The authors state, ``In a weak field of about $10^{-4}$~T, the Cs-$^{129}$Xe enhancement factor is reported to be $10^4$ to $10^5$," but no such measurement has ever been made. The citations \cite{Driehuys1995,Zhou2004} in support of this statement (authors' Refs.~[22] and [23]) are to papers that have nothing to do with measuring $\kappa$; rather, the ``enhancement" discussed in those works refers to the increase in $^{129}$Xe polarization due to SEOP. Driehuys, et al.~\cite{Driehuys1995} showed a dependence of $^{129}$Xe {\em surface relaxation} on magnetic field (as the title of that manuscript suggests), which is completely unrelated to $\kappa$.

The enhancement factor $\kappa$ has to do with the shape of the relevant Cs-Xe interatomic potentials and the perturbed Cs electron wave function \cite{Walker1989,Schaefer1989}; the Cs-Xe molecular lifetime has to do with the gas kinetics of neutral atoms. Neither parameter has any intrinsic dependence on magnetic field. The authors' statement that ``the lifetime of the van der Waals (vdW) molecule is shortened in a strong magnetic field," is thus also incorrect.  Molecular lifetime is inversely proportional to gas {\em density} (at fixed gas composition), which means that gas density can also affect $\kappa$. The vapor cell used in \cite{Xu2021} has a total pressure of $\approx 120$~Torr, which could only {\em lower} $\kappa$ from it's value in the high-pressure limit (referred to as $\kappa_0$), and even then by no more than about 10\% \cite{Schaefer1989,Walker1989}.

What is true (and perhaps what the authors are trying to point to) is that the vdW-molecular contribution to the spin-exchange rate can change as a function of applied magnetic field $B_0$; however, the relevant comparisons are not discussed anywhere in the paper. The relevant fields are the alkali-metal hyperfine field and the molecular spin-rotation field  \cite{Appelt1998,Happer1984}. There is a muddled discussion of the regimes (binary vs.\ molecular) in which SEOP of the heavier noble gases has been studied, further complicated by the failure to distinguish between the ''spin-exchange cross section," which should encompass {\em all} SEOP mechanisms (binary and molecular), and the separate binary and molecular contributions, each of which has its own cross section. In any case, the discussion of the magnetic-field dependence of the rate coefficients associated with SEOP is not terribly relevant to the nature or measurement of $\kappa$, which has no magnetic-field dependence.

According to the authors' Eq.~(3), there are three quantities that they must measure independently with reasonable accuracy to obtain an accurate measurement of $\kappa$: the effective magnetic field produced by nuclear spins $B^{\rm eff}_z$, the nuclear spin polarization $P_n$, and the density $n$ of the relevant atoms. The measurements of these quantities appear to be reasonably executed, although accurate measurements at the cell-filling step of such small densities of Xe can be difficult, and the authors do not discuss the details of their procedure. The authors' measurement of $P_n$ requires an accurate measurement of the Cs polarization $P^e$, which, unlike $P_n$, can be quite inhomogeneous across the cell volume. Most importantly, the authors' analysis is done assuming that their Eq.~(3) applies exclusively to $^{129}$Xe.  We first note that the dependence of $\kappa$ on atomic mass of the Xe isotopes is {\em negligible}, as the fractional difference in masses is so small as to have almost no effect on the collision trajectories. The authors incorrectly argue that, ``Since the enhancement factor of $^{131}$Xe is much smaller than $^{129}$Xe, it is reasonable to neglect the contribution to the effective magnetic field from the $^{131}$Xe in this estimation." Since their cell has equal amounts of $^{129}$Xe and $^{131}$Xe, and since the magnetic moment of $^{129}$Xe is only about 10\% larger than that of $^{131}$Xe, the only way that $^{131}$Xe can be ignored is if its SEOP-generated polarization is much smaller than that of $^{129}$Xe. The $^{131}$Xe polarization is probably smaller (due to faster non-SEOP-driven $T_1$-relaxation) but to what degree? It does not appear that the $^{131}$Xe polarization was measured, and ignoring it in this case could lead to a significant miscalculation of the effective field (and thus of $\kappa$).

Taken together, the concerns outlined here cast significant doubt on the accuracy of the measurement in \cite{Xu2021}. Given that previous measurements and theoretical estimates are more than an order of magnitude smaller than the authors' result, there is little reason to accept or consider this result in its current form in the context of future relevant work on Cs-Xe SEOP and its applications.

\end{document}